\documentclass[aps,prl,twocolumn,showpacs,superscriptaddress]{revtex4}  
\usepackage{graphicx}  
\usepackage{dcolumn}   
\usepackage{bm}        
\usepackage{amssymb}   
\usepackage{multirow}
\usepackage{units}

\newcommand{\etal}{{\it et al.}}
\newcommand{\pot}{\ensuremath{3.14\times10^{20}} protons-on-target}
\newcommand{\mcnnpred}{\ensuremath{22\pm 5 {\rm (stat.)} \pm 3 {\rm (syst.)}}}
\newcommand{\annpred}{\ensuremath{27\pm 5 {\rm (stat.)} \pm 2 {\rm (syst.)}}}
\newcommand{\mcnnobs}{\ensuremath{28}}
\newcommand{\annobs}{\ensuremath{35}}

\newcommand{\delmsq}[1]{\ensuremath{\Delta m^2_{ #1 }}}
\newcommand{\dmsq}[1]{\delmsq{ #1 }}
\newcommand{\sinsq}[1]{\ensuremath{\sin^{2}\left(2\theta_{ #1 }\right)}}

\def\numunue{$\nu_\mu \rightarrow \nu_e$}

\def\numunutau{$\nu_\mu \rightarrow \nu_{\tau}$}

\newcommand{\numu}{\mbox{$\nu_{\mu}$}}                   
\newcommand{\nue}{\mbox{$\nu_{e}$}}                      
\newcommand{\nutau}{\mbox{$\nu_{\tau}$}}                 
\newcommand{\anue}{\mbox{$\overline{\nu}_{e}$}}          
\newcommand{\anumu}{\mbox{$\overline{\nu}_{\mu}$}}       

\newcommand{\piz}{\mbox{$\pi^{0}$}}                      

\begin{document}
\pacs{14.60.Pq, 14.60.Lm, 29.27.-a}

\title{Search for muon-neutrino to electron-neutrino transitions in MINOS}






\newcommand{\Cambridge}{Cavendish Laboratory, University of Cambridge, Madingley Road, Cambridge CB3 0HE, United Kingdom}
\newcommand{\FNAL}{Fermi National Accelerator Laboratory, Batavia, Illinois 60510, USA}
\newcommand{\RAL}{Rutherford Appleton Laboratory, Science and Technology Facilities Council, OX11 0QX, United Kingdom}
\newcommand{\UCL}{Department of Physics and Astronomy, University College London, Gower Street, London WC1E 6BT, United Kingdom}
\newcommand{\Caltech}{Lauritsen Laboratory, California Institute of Technology, Pasadena, California 91125, USA}
\newcommand{\ANL}{Argonne National Laboratory, Argonne, Illinois 60439, USA}
\newcommand{\Athens}{Department of Physics, University of Athens, GR-15771 Athens, Greece}
\newcommand{\NTUAthens}{Department of Physics, National Tech. University of Athens, GR-15780 Athens, Greece}
\newcommand{\Benedictine}{Physics Department, Benedictine University, Lisle, Illinois 60532, USA}
\newcommand{\BNL}{Brookhaven National Laboratory, Upton, New York 11973, USA}
\newcommand{\CdF}{APC -- Universit\'{e} Paris 7 Denis Diderot, 10, rue Alice Domon et L\'{e}onie Duquet, F-75205 Paris Cedex 13, France}
\newcommand{\Cleveland}{Cleveland Clinic, Cleveland, Ohio 44195, USA}
\newcommand{\Delhi}{Department of Physics and Astrophysics, University of Delhi, Delhi 110007, India}
\newcommand{\GEHealth}{GE Healthcare, Florence South Carolina 29501, USA}
\newcommand{\Harvard}{Department of Physics, Harvard University, Cambridge, Massachusetts 02138, USA}
\newcommand{\HolyCross}{Holy Cross College, Notre Dame, Indiana 46556, USA}
\newcommand{\IIT}{Physics Division, Illinois Institute of Technology, Chicago, Illinois 60616, USA}
\newcommand{\Iowa}{Department of Physics and Astronomy, Iowa State University, Ames, Iowa 50011, USA}
\newcommand{\Indiana}{Indiana University, Bloomington, Indiana 47405, USA}
\newcommand{\ITEP}{High Energy Experimental Physics Department, ITEP, B. Cheremushkinskaya, 25, 117218 Moscow, Russia}
\newcommand{\JMU}{Physics Department, James Madison University, Harrisonburg, Virginia 22807, USA}
\newcommand{\LASL}{Nuclear Nonproliferation Division, Threat Reduction Directorate, Los Alamos National Laboratory, Los Alamos, New Mexico 87545, USA}
\newcommand{\Lebedev}{Nuclear Physics Department, Lebedev Physical Institute, Leninsky Prospect 53, 119991 Moscow, Russia}
\newcommand{\LLL}{Lawrence Livermore National Laboratory, Livermore, California 94550, USA}
\newcommand{\MIT}{Lincoln Laboratory, Massachusetts Institute of Technology, Lexington, Massachusetts 02420, USA}
\newcommand{\Minnesota}{University of Minnesota, Minneapolis, Minnesota 55455, USA}
\newcommand{\Crookston}{Math, Science and Technology Department, University of Minnesota -- Crookston, Crookston, Minnesota 56716, USA}
\newcommand{\Duluth}{Department of Physics, University of Minnesota -- Duluth, Duluth, Minnesota 55812, USA}
\newcommand{\Otterbein}{Otterbein College, Westerville, Ohio 43081, USA}
\newcommand{\Oxford}{Subdepartment of Particle Physics, University of Oxford, Oxford OX1 3RH, United Kingdom}
\newcommand{\Pittsburgh}{Department of Physics and Astronomy, University of Pittsburgh, Pittsburgh, Pennsylvania 15260, USA}
\newcommand{\IHEP}{Institute for High Energy Physics, Protvino, Moscow Region RU-140284, Russia}
\newcommand{\RoyalH}{Physics Department, Royal Holloway, University of London, Egham, Surrey, TW20 0EX, United Kingdom}
\newcommand{\Carolina}{Department of Physics and Astronomy, University of South Carolina, Columbia, South Carolina 29208, USA}
\newcommand{\SLAC}{Stanford Linear Accelerator Center, Stanford, California 94309, USA}
\newcommand{\Stanford}{Department of Physics, Stanford University, Stanford, California 94305, USA}
\newcommand{\StJohnFisher}{Physics Department, St. John Fisher College, Rochester, New York 14618 USA}
\newcommand{\Sussex}{Department of Physics and Astronomy, University of Sussex, Falmer, Brighton BN1 9QH, United Kingdom}
\newcommand{\TexasAM}{Physics Department, Texas A\&M University, College Station, Texas 77843, USA}
\newcommand{\Texas}{Department of Physics, University of Texas at Austin, 1 University Station C1600, Austin, Texas 78712, USA}
\newcommand{\TechX}{Tech-X Corporation, Boulder, Colorado 80303, USA}
\newcommand{\Tufts}{Physics Department, Tufts University, Medford, Massachusetts 02155, USA}
\newcommand{\UNICAMP}{Universidade Estadual de Campinas, IFGW-UNICAMP, CP 6165, 13083-970, Campinas, SP, Brazil}
\newcommand{\USP}{Instituto de F\'{i}sica, Universidade de S\~{a}o Paulo,  CP 66318, 05315-970, S\~{a}o Paulo, SP, Brazil}
\newcommand{\Warsaw}{Department of Physics, University of Warsaw, Ho\.{z}a 69, PL-00-681 Warsaw, Poland}
\newcommand{\Washington}{Physics Department, Western Washington University, Bellingham, Washington 98225, USA}
\newcommand{\WandM}{Department of Physics, College of William \& Mary, Williamsburg, Virginia 23187, USA}
\newcommand{\Wisconsin}{Physics Department, University of Wisconsin, Madison, Wisconsin 53706, USA}
\newcommand{\deceased}{Deceased.}

\affiliation{\ANL}
\affiliation{\Athens}
\affiliation{\Benedictine}
\affiliation{\BNL}
\affiliation{\Caltech}
\affiliation{\Cambridge}
\affiliation{\UNICAMP}
\affiliation{\CdF}
\affiliation{\FNAL}
\affiliation{\Harvard}
\affiliation{\HolyCross}
\affiliation{\IIT}
\affiliation{\Indiana}
\affiliation{\Iowa}
\affiliation{\Lebedev}
\affiliation{\LLL}
\affiliation{\UCL}
\affiliation{\Minnesota}
\affiliation{\Duluth}
\affiliation{\Otterbein}
\affiliation{\Oxford}
\affiliation{\Pittsburgh}
\affiliation{\RAL}
\affiliation{\USP}
\affiliation{\Carolina}
\affiliation{\Stanford}
\affiliation{\Sussex}
\affiliation{\TexasAM}
\affiliation{\Texas}
\affiliation{\Tufts}
\affiliation{\Warsaw}
\affiliation{\Washington}
\affiliation{\WandM}
\affiliation{\Wisconsin}

\author{P.~Adamson}
\affiliation{\FNAL}

\author{C.~Andreopoulos}
\affiliation{\RAL}

\author{K.~E.~Arms}
\affiliation{\Minnesota}

\author{R.~Armstrong}
\affiliation{\Indiana}

\author{D.~J.~Auty}
\affiliation{\Sussex}


\author{D.~S.~Ayres}
\affiliation{\ANL}

\author{C.~Backhouse}
\affiliation{\Oxford}



\author{P.~D.~Barnes~Jr.}
\affiliation{\LLL}

\author{G.~Barr}
\affiliation{\Oxford}

\author{W.~L.~Barrett}
\affiliation{\Washington}


% ...

\collaboration{The MINOS Collaboration}
\noaffiliation




\date{\today}

\begin{abstract}
This letter reports on a search for $\numu{}\rightarrow\nue{}$ transitions by the MINOS experiment based on a \unit[$3.14\times10^{20}$]{protons-on-target} exposure in the Fermilab NuMI beam.  We observe \annobs{} events in the Far Detector with a background of \annpred{} events predicted by the measurements in the Near Detector. If interpreted in terms of $\numu{}\rightarrow\nue{}$ oscillations, this 1.5\,$\sigma$ excess of events is consistent with $\sin^{2}(2\theta_{13})$ comparable to the CHOOZ limit when $|\delmsq{}|$=\unit[2.43$\times 10^{-3}$] {${\rm eV^{2}}$} and \sinsq{23}=1.0 are assumed.

\end{abstract}

\maketitle

Several experiments have provided compelling evidence for muon neutrino disappearance as a function of neutrino energy and distance traveled~\cite{ref:osc1,ref:osc3,ref:osc4,ref:osc5,ref:imb, ref:kamioka,ref:minos08}.  These observations support the description of neutrinos in two distinct mass and flavor bases, related by the $3\times 3$ neutrino mixing matrix~\cite{ref:PNMS}. The MINOS experiment provides the most precise measurement of the atmospheric mass splitting, $|\delmsq{}|= $ \unit[(2.43$\pm 0.13 )\times 10^{-3}$]{${\rm eV^{2}}$} ~\cite{ref:dm2note, ref:minos08}.  At this mass scale, the dominant oscillation channel is expected to be \numunutau{}, but sub-dominant \numunue{} transitions are not excluded~\cite{ref:tausuperk}.  Observation of \nue\ appearance would imply a non-zero value of $\theta_{13}$, opening the possibility of observing CP violation in the leptonic sector.  The current best experimental limit~\cite{ref:chooz}, implies $\sin^{2}(2\theta_{13})<0.15$ at the 90\% confidence level (C.L.) for the MINOS $|\delmsq{}|$ value. In addition to these parameters, the probability of electron neutrino transitions in MINOS depends on  $\sin^{2}(2\theta_{23})$, the CP violation parameter, $\delta_{CP}$, and the sign of $\delmsq{}$. Two other experiments have given limits with less sensitivity \cite{ref:k2k, ref:paloalto}; MINOS is the first experiment to probe $\sin^{2}(2\theta_{13})$ with sensitivity comparable to the CHOOZ limit at $|\delmsq{}|$=\unit[2.43$\times 10^{-3}$] {${\rm eV^{2}}$} and \sinsq{23}=1.0. 

Neutrino interactions from a beam produced by the Fermilab NuMI facility~\cite{ref:beam} are recorded at the MINOS Near (ND) and Far (FD) Detectors, located at \unit[1]{km} and \unit[735]{km}, respectively, from the production target. High statistics data from the ND establish the properties of the beam before oscillations. Observation of additional \nue{} interactions in the FD relative to the ND provides evidence of $\numu{}\rightarrow\nue{}$  oscillation. The two detectors are of similar design to reduce systematic uncertainties from the physics of neutrino interactions, the neutrino flux, and detector response~\cite{ref:minos08, ref:minosnim, ref:caldetrel}. The detectors are magnetized, tracking calorimeters composed of planes each with layers of \unit[2.54]{cm} thick steel and \unit[1.0]{cm} thick scintillator ($1.4$ radiation lengths per plane). The scintillator layer is composed of \unit[4.1]{cm} wide strips ($1.1$ Moli\`ere radii). 

The beam is comprised of 98.7\% \numu{}+\anumu{} and 1.3\% \nue{}+\anue{}. The latter originate from decays of muons produced in pion decays and from kaon decays. The \nue{} flux below \unit[8]{GeV} is largely from muon decay and is well constrained by the measured \numu{} flux~\cite{ref:minos08, ref:PRD}. The present analysis is based on an integrated exposure of $3.14 \times 10^{20}$ protons delivered to the NuMI target.

The search for electron neutrino appearance relies on identifying charged current (CC) $\nue{}+{\rm Fe}\rightarrow e+X$ interactions that produce an energetic electron~\cite{ref:nubarnote}. This electron initiates an electromagnetic cascade and deposits its energy in a relatively narrow and short region in the MINOS calorimeter. Additional calorimeter activity is produced by the breakup of the recoil nucleus ($X$). Other neutrino scattering processes can produce similar event topologies in the MINOS detector. These include neutral current (NC) $\nu+{\rm Fe}\rightarrow \nu+X$ interactions and \numu{}-CC interactions with low energy muons, both having hadronic showers with an electromagnetic component arising from \piz\ decays.  Less significant backgrounds arise from intrinsic beam \nue{}-CC interactions, \nutau{}-CC interactions from oscillations, and cosmogenic backgrounds.

We select events with reconstructed energy between 1 and 8 GeV, encompassing the maximum of the \numunue\ oscillation probability.  The lower limit mainly removes NC events, while the higher limit removes beam \nue{}-CC events from kaon decays.  Additionally, events are selected to be in time with the accelerator beam pulse, and directional requirements are applied to limit background from cosmogenic sources to less than 0.5 events (90\% C.L.). Events are required to have a reconstructed shower and at least 5 contiguous planes each with energy deposition above \unit[1]{MeV}.  Events with tracks longer than 25 planes are rejected.  Monte Carlo (MC) simulations indicate that these cuts improve the signal-to-background ratio from 1:55 to 1:12, assuming \sinsq{13}$=0.15$.

Further enrichment of the \nue{}-CC selected sample is achieved using a method based on an artificial neural network (ANN) with 11 input variables characterizing the longitudinal and transverse energy deposition in the calorimeter that separate the signal \nue{}-CC events from NC and \numu{}-CC background~\cite{ref:TJthesis}. The acceptance threshold is determined by maximizing the ratio of the accepted signal to the expected statistical and systematic uncertainty of the background. With these selection criteria, and assuming \sinsq{13}$=0.15$, this method gives a 1:4 signal-to-background ratio. 

\begin{figure}
\begin{center}
\includegraphics[viewport=50 50  567 448, keepaspectratio,width= 0.48 \textwidth, clip=true]{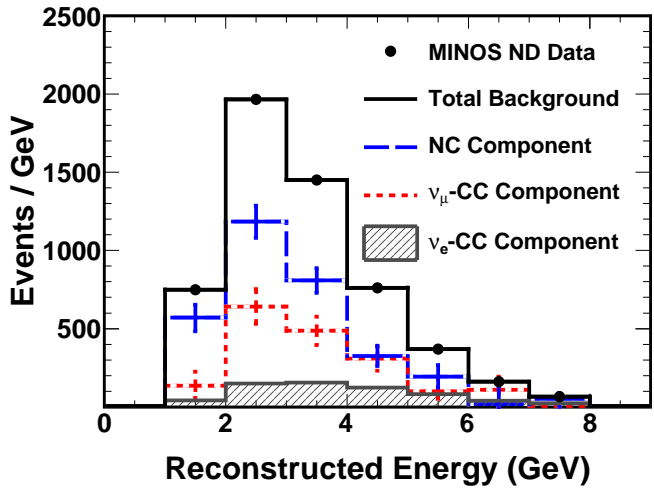}

\end{center}
\caption{Reconstructed ND energy spectra of the  \nue{}-CC selected backgrounds from NC (dashed) and \numu-CC (dotted) interactions as obtained from the horn-off method. The shaded histogram shows the beam \nue-CC component from the simulation. The solid histogram corresponds to the total of these three components which are constrained to agree with the data points. The uncertainties shown on the data are statistical and are not visible on this scale; uncertainties on the components are systematic. }
\label{fig:nddata}
\end{figure}

The Monte Carlo simulation of the beam line and detectors is based on {\tt GEANT3}~\cite{ref:geant} and the hadron production yields from the target are calculated by {\tt FLUKA}~\cite{ref:fluka}. The calculated neutrino flux is adjusted to agree with the ND \numu{}-CC data~\cite{ref:PRD}. Neutrino interactions and further re-interactions of the resulting hadrons within the nucleus are simulated with {\tt NEUGEN3}~\cite{ref:neugen}. Uncertainties in the composition and kinematic distribution of the particles that emerge from the nucleus can be large, but these and many other uncertainties mostly cancel when comparing neutrino interactions between the ND and FD.  

Events selected by the ANN in the ND predict the number of background events expected in the FD and reduce reliance on the simulation. Considerations of oscillations and beam line geometry require that each background component, $\numu$-CC, NC, and beam $\nue$-CC, be treated separately in the prediction of the FD backgrounds.  The first two background components are determined using an NC-enriched data sample recorded with the NuMI magnetic horns turned off.  In this configuration the pions are not focused; the low-energy peak of the neutrino energy distribution disappears, leaving an event sample dominated by NC events from higher energy neutrino interactions.  These data are used in conjunction with the standard beam configuration data, and the simulated ratios of the horn-on to horn-off rates for each component, to extract the individual NC and \numu{}-CC background spectra. The smaller beam $\nue$-CC component is calculated from the MC using  $\numu$-CC events observed in the ND. 

Figure~\ref{fig:nddata} shows the data in the ND and the derived NC,  \numu{}-CC, and beam \nue{}-CC backgrounds.  The ND background is (57$\pm$5)\% NC, (32$\pm$7)\% \numu{}-CC and (11$\pm$3)\% beam \nue{}-CC events. Systematic errors on the components arise from uncertainties in the beam flux, cross-section and selection efficiency. The errors on the NC and \numu{}-CC components are derived from the data and are correlated due to the constraint that the background must add to the observed ND event rate.  The uncertainty on the beam \nue{}-CC comes from the \numu{}-CC events observed in the ND~\cite{ref:PRD}.

As a crosscheck, a second technique to study the ND background sample uses an independent sample of showers from \numu-CC events selected with long tracks~\cite{ref:minos08}.  The hits associated with the muon track are removed from the event, and the remnant showers are subsequently analyzed as a sample of NC-like events~\cite{ref:nomad, ref:JBthesis}. The procedure is applied to the data and MC, and the \nue{}-CC selections are applied to both.  Differences between the muon-removed data and muon-removed MC samples are used to adjust the predicted NC background. As in the first method, the beam \nue{}-CC background is taken from the MC, and the remainder of the observed ND background are classified as \numu-CC events. The background components calculated from the muon-removed sample agree with those obtained from the horn-off method~\cite{ref:AHthesis}. 

After decomposition of the ND energy spectrum into background components, each of these spectra is multiplied by the Far to Near energy spectrum ratio from the simulation for that component, providing a prediction of the FD spectrum. The simulation takes into account differences in the spectrum of events at the ND and FD due to beam line geometry as well as possible differences in detector calibrations and event shape.  Oscillations are included when predicting the \numu{}-CC component.  We expect 26.6 background events, of which 18.2 are NC, 5.1 are \numu-CC, 2.2  are beam \nue{} and 1.1 are \nutau{}~\cite{ref:oscpars}.

To estimate the efficiency for selecting \nue{}-CC events, we use the muon-removed events from data and MC, then embed a simulated electron of the same momentum as the removed muon.  Test beam measurements~\cite{ref:caldet} indicate that the selection efficiency of single electrons agrees with the simulation to within 2.6\%.  The \nue{}-CC selection efficiency obtained from the data agrees with the selection efficiency obtained from the MC to within 0.3\%.  We estimate our efficiency for selecting \nue{}-CC events to be (41.4$\pm$1.5)\%~\cite{ref:JBthesis}. 

Systematic uncertainties are evaluated by generating modified MC samples and by quantifying the change in the number of predicted background events in the FD. Table~\ref{tab:systematics} shows that the dominant uncertainties arise from Far/Near differences:  relative energy scale calibration differences  (a),  details of the modeling of the photomultiplier tube (PMT) gains (b) and crosstalk (c), and relative event rate normalization (d).  Other uncertainties resulting from neutrino interaction physics, shower hadronization, intranuclear rescattering, and absolute energy scale uncertainties (e) affect the events in both detectors in a similar manner and mostly cancel in the extrapolation. The individual systematic uncertainties are added in quadrature along with the uncertainty arising from the background decomposition in the ND to give an overall systematic uncertainty of 7.3\% on the expected number of background events selected in the FD.  The expected statistical uncertainty is 19\%.

\begin{table}
\begin{tabular}{l l c}
\hline \hline 
\multirow{2}{*}{Uncertainty source} & & Uncertainty on\\
 & &  background events \\
\hline 
Far/Near ratio: & &  6.4\% \\
\ \ (a) Relative Energy Scale  &3.1\%& \\
\ \ (b) PMT Gains &2.7\% & \\
\ \ (c) PMT Crosstalk & 2.2\%& \\
\ \ (d) Relative Event Rate  & 2.4\%& \\
\ \ (e) All Others & 3.7\%& \\

Horn-off (systematic) & & 2.7\% \\
Horn-off (statistical) & & 2.3\% \\

\hline
Total Systematic Uncertainty && 7.3\% \\
\hline \hline

\end{tabular}
\caption{Systematic uncertainty in the total number of background events in the Far Detector.}
\label{tab:systematics}
\end{table}

The prediction and uncertainty of the backgrounds in the FD are established before examining the FD data.  Additionally, the independent and signal-free muon-removed FD data sample is examined.  In that sample, we observe 39 events, with an expectation of $29\pm5{\rm (stat.)}\pm2{\rm(syst.)}$. The selected events were investigated and no evidence of abnormalities was found.

\begin{figure}
\begin{center}
\includegraphics[viewport=50 50  567 427, keepaspectratio,width= 0.45 \textwidth, clip=true]{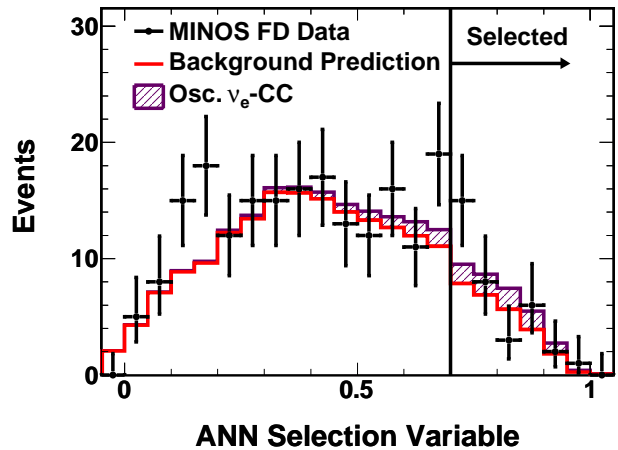}
\end{center}
\caption{Distribution of the ANN selection variable for events in the FD.  Black points show data with statistical error bars. The non-shaded histogram shows the background expectation. The shaded region shows the additional  \nue{}-CC events required to explain the observed excess with the oscillation hypothesis.}
\label{fig:fdpid}
\end{figure}

\begin{figure}
\begin{center}

\includegraphics[viewport=50 50  567 415, keepaspectratio,width= 0.45 \textwidth, clip=true]{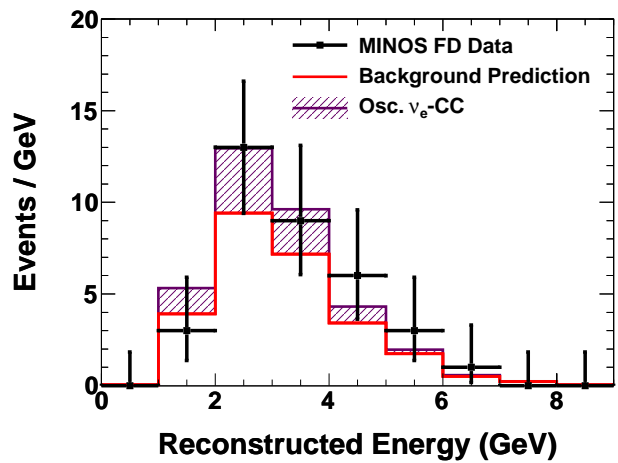}

\end{center}
\caption{Reconstructed energy distribution of the \nue{}-CC selected events in the FD, with the exception of the energy cut.  No data events are found below \unit[1]{GeV} or above \unit[8]{GeV}, consistent with expectation. Black points show data with statistical error bars.  The non-shaded histogram shows the background expectation. The shaded region shows the additional  \nue{}-CC events required to explain the observed excess with the oscillation hypothesis.}
\label{fig:fdspectrum}
\end{figure}

Figure~\ref{fig:fdpid} shows the FD data as a function of the ANN selection variable.  The signal acceptance threshold was optimized prior to examination of the FD data to be $0.7$.  We observe \annobs{} events in the signal region with a background expectation of \annpred{}.  In the region of the selection variable well below the acceptance threshold ($<0.55$), we observe 146 events, compared to a pure background expectation of $132\pm12{\rm (stat.)}\pm8{\rm(syst.)}$.  The observed energy spectrum for the events in the signal region is shown in Figure~\ref{fig:fdspectrum}.

A second selection method, Library Event Matching, is used as a crosscheck.  In this technique, each candidate is compared to a large library of simulated \nue-CC and NC events~\cite{ref:POthesis}. This method gives a better background rejection than the ANN algorithm, but with increased sensitivity to some uncertainties. As in the ANN method, we observe a small excess ($<2\sigma$) for the muon-removed sample and in the region below the selection cut. In the signal region, we observe \mcnnobs{} selected events, with a background expectation of \mcnnpred{}; these results are consistent with the ANN selection.

\begin{figure}
\begin{center}
\includegraphics[viewport=25 25  467 650, keepaspectratio,width= 0.45 \textwidth, clip=true]{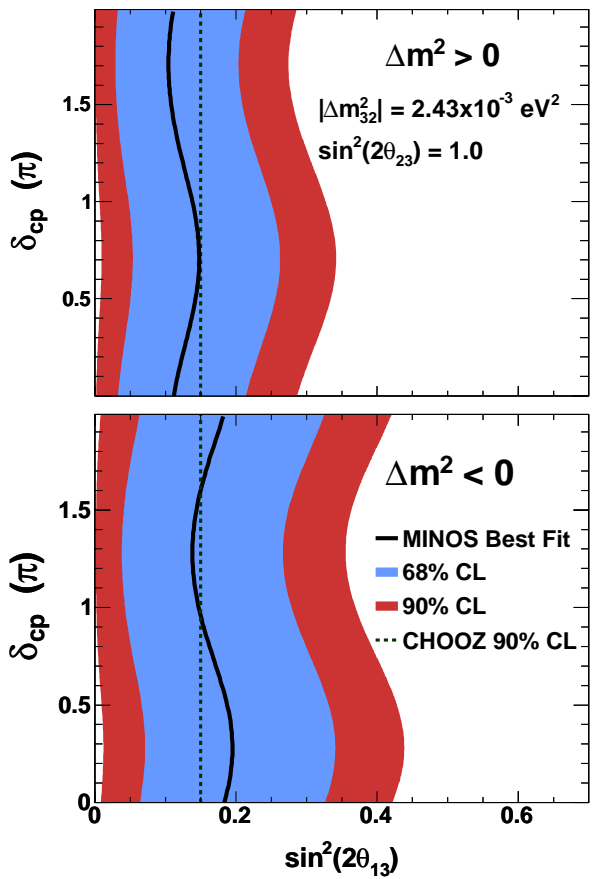}
\end{center}
\caption{Values of  \sinsq{13} and $\delta_{CP}$ that produce a number of events consistent with the observation for the normal hierarchy (top) and inverted hierarchy (bottom).  Black lines show those values that best represent our data. Red (blue) regions show the 90\% (68\%) C.L. intervals.}
\label{fig:sens}
\end{figure}

Figure~\ref{fig:sens} shows the values of \sinsq{13} and $\delta_{CP}$ that give an excess of events consistent with our observation from the ANN selection.  The oscillation probability is computed using a full 3-flavor neutrino mixing framework that includes matter effects~\cite{ref:3flav}, which introduces a dependence on the neutrino mass hierarchy (the sign of $\delmsq{}$). The MINOS best fit values of  $|\delmsq{}|$=\unit[2.43$\times 10^{-3}$] {${\rm eV^{2}}$} and  \sinsq{23}=1.0 are used as constants in the calculation~\cite{ref:sinsqft}. Statistical fluctuations (Poisson) and systematic effects (Gaussian) are incorporated via the Feldman-Cousins approach~\cite{ref:FC} which determines the confidence intervals. 

In conclusion, we report the first results of a search for \nue{} appearance by the MINOS experiment. The \annobs{} events in the Far Detector after \nue{} selection for \pot{} are 1.5\,$\sigma$ higher than the background expectation of \annpred{}.  Assuming  $|\delmsq{}|$=\unit[2.43$\times 10^{-3}$] {${\rm eV^{2}}$} and  \sinsq{23}=1.0, the best fit for the normal hierarchy is just below the CHOOZ~\cite{ref:chooz} limit for all values of $\delta_{CP}$. 

This work was supported by the US DOE; the UK STFC; the US NSF; the State and University of Minnesota; the University of Athens, Greece; and Brazil's FAPESP and CNPq.  We are grateful to the Minnesota Department of Natural Resources, the crew of the Soudan Underground Laboratory, and the staff of Fermilab for their contribution to this effort.


\begin{thebibliography}{99}
\bibitem{ref:nubarnote} {The experiment does not distinguish  \nue\ and \anue; the dominant contribution is expected from $\numu{}\rightarrow\nue{}$ oscillations.}
\bibitem{ref:sinsqft}
Electron neutrino appearance in MINOS is proportional to $2\sin^2(\theta_{23})\sin^2(2\theta_{13})$.
\end{thebibliography}
\end{document}